\documentclass[letterpaper, 10 pt, conference]{ieeeconf}  

\IEEEoverridecommandlockouts                              

\usepackage[noadjust, nobreak]{cite}

\usepackage{graphicx}
\graphicspath{{./graphics/}}
\usepackage[utf8]{inputenc}
\usepackage{multirow}

\usepackage{xcolor}

\usepackage{amsmath}
\usepackage[noadjust]{cite}

\usepackage{balance}

\title{\LARGE \bf The Atlas of Lane Changes: \linebreak Investigating Location-dependent Lane Change Behaviors \linebreak Using Measurement Data from a Customer Fleet
}

\author{Florian~Wirthmüller\textsuperscript{\includegraphics[scale=0.4]{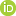}},
	Jochen~Hipp\textsuperscript{\includegraphics[scale=0.4]{orcidlogo.png}},
        Christian~Reichenbächer\textsuperscript{\includegraphics[scale=0.4]{orcidlogo.png}}
        and~Manfred~Reichert\textsuperscript{\includegraphics[scale=0.4]{orcidlogo.png}}%
\thanks{F.~Wirthmüller, J.~Hipp and C.~Reichenbächer are with Mercedes-Benz AG, Böblingen, Germany, E-Mail: \{first\_name.last\_name\}@daimler.com.}
\thanks{F.~Wirthmüller and M.~Reichert are with the Institute of Databases and Information Systems (DBIS) at Ulm University, Ulm, Germany,\newline E-Mail: \{first\_name.last\_name\}@uni-ulm.de}
\thanks{C.~Reichenbächer is with the Wilhelm-Schickard-Institute for Informatics at Eberhard Karls University Tübingen, Tübingen, Germany.}
\thanks{ORCID (ordered as authors above): \newline \href{https://orcid.org/0000-0002-9732-2561}{https://orcid.org/0000-0002-9732-2561};\newline \href{https://orcid.org/0000-0002-9037-9899}{https://orcid.org/0000-0002-9037-9899};\newline \href{https://orcid.org/0000-0002-0907-3287}{https://orcid.org/0000-0002-0907-3287};\newline \href{https://orcid.org/0000-0003-2536-4153}{https://orcid.org/0000-0003-2536-4153}}\thanks{\copyright~2021 IEEE. Personal use of this material is permitted. Permission from IEEE must be obtained for all other uses, in any current or future media, including reprinting/republishing this material for advertising or promotional purposes, creating new collective works, for resale or redistribution to servers or lists, or reuse of any copyrighted component of this work in other works.}
}

\usepackage{url}
\usepackage{hyperref}

\begin{document}
\IEEEoverridecommandlockouts
\pubid{\copyright~2021 IEEE}

\maketitle
\pagestyle{empty}


\begin{abstract}

The prediction of surrounding traffic participants behavior is a crucial and challenging task for driver assistance and autonomous driving systems. Today's approaches mainly focus on modeling dynamic aspects of the traffic situation and try to predict traffic participants behavior based on this. In this article we take a first step towards extending this common practice by calculating location-specific a-priori lane change probabilities. The idea behind this is straight forward: The driving behavior of humans may vary in exactly the same traffic situation depending on the respective location. E.\,g. drivers may ask themselves: Should I pass the truck in front of me immediately or should I wait until reaching the less curvy part of my route lying only a few kilometers ahead? Although, such information is far away from allowing behavior prediction on its own, it is obvious that today’s approaches will greatly benefit when incorporating such location-specific a-priori probabilities into their predictions. For example, our investigations show that highway interchanges tend to enhance driver's motivation to perform lane changes, whereas curves seem to have lane change-dampening effects. Nevertheless, the investigation of all considered local conditions shows that superposition of various effects can lead to unexpected probabilities at some locations. We thus suggest dynamically constructing and maintaining a lane change probability map based on customer fleet data in order to support onboard prediction systems with additional information. For deriving reliable lane change probabilities a broad customer fleet is the key to success.
\end{abstract}

\section{INTRODUCTION}

\begin{figure}[t!]
\centering\includegraphics[width=0.46\textwidth]{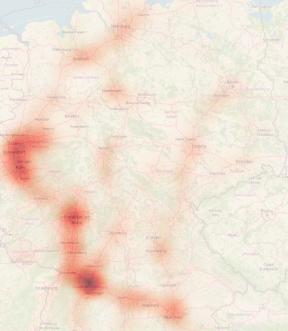}\caption{Heatmap of lane changes detected in the collected measurements on German highways. Basemap: OpenStreetMap \cite{OpenStreetMap}}\label{fig:lcmap}
\end{figure}

In general, experienced drivers are aware that they behave differently when facing the same traffic situation depending on the respective location. In this context, various local conditions, as e.\,g. the curviness, the width of the road and the roadside structures or the placement of the signage, play an important role. 


In order to navigate safely and comfortably for the passengers through any traffic situation automated vehicles need to predict surrounding vehicle's behaviors. Although this is already a challenging task, the above described effect makes it even more challenging. In order to support such predictions, this article aims to analyze which local conditions are the most important and to which extent they influence the driving behavior. In our research, we focus on the lane change behavior in highway scenarios.

\pubidadjcol

For our analyses, we collect measurement data from a fleet of customer vehicles. Note that most automotive software functions are developed and tested based on data collected with dedicated test vehicles, so far. This, however, results in several limitations. For instance, test drivers, knowing their route very well tend to drive in a special manner. Data collected from real customers therefore might be more representative. Besides, considering customer data offers the chance to cover much more miles in less time.

After collecting the data, they are processed and spatially aggregated using digital maps. As a result we obtain a map of lane change probabilities. Together with other map attributes this enables downstreamed analyses. 


In detail, we make the following contributions: 

\begin{enumerate}
\item A procedure to construct a lane change probability map
\item A thorough study and discussion of the impact of three exemplary location-dependent conditions (interchanges, curvature, slope) on the lane change behavior
\end{enumerate}

The remainder of this article is structured as follows: \autoref{sec:rel_work} discusses related work. \autoref{sec:prep} outlines the preparation of the map and measurement data. Subsequently, \autoref{sec:results} presents our empiric investigations and discusses the results obtained. Finally, \autoref{sec:conclusion} summarizes our contribution and provides an outlook on future work.

\section{RELATED WORK}\label{sec:rel_work}

A general overview on motion prediction approaches and especially deep learning-based ones is given in \cite{lefevre2014, mozaffari2020deep}. Besides characterizing and categorizing approaches, \cite{mozaffari2020deep} shows typical strengths and weaknesses of each group. Among other things, it is mentioned that environment conditions, such as, road geometry and traffic rules can have significant impacts on the driving behavior. Most shown approaches are, however, not able to handle such impacts due to their input representation. For the latter, most approaches solely rely on the motion history of the vehicle to be predicted as well as its surrounding vehicles history. Other approaches building up on raw sensor data or birds eye views are on the other hand computationally expensive. This makes such approaches hard to implement on onboard hardware with limited resources.

\cite{wirthmueller2020} agrees with \cite{mozaffari2020deep} about the importance of external conditions for the driving behavior. The authors explicitly mention weather, daytime and traffic density. Subsequently, the latter's influence on the prediction capability of a known motion prediction system rather than on the driving behavior itself is studied. The article confirms the influence of varying traffic densities and concludes that such impacts need to be studied more in detail and integrated in motion prediction approaches.

\cite{wirthmuller2020fleet} describes the development of an architecture concept aiming to enhance behavior predictions through massive online learning using customer fleets. The general idea here is to learn several context specific prediction models and provide them to all vehicles.

\cite{imanishi2020model} is also tackling the motion prediction problem in diverse environments. In order to that, all vehicles of the fleet are permanently connected to a cloud server. The vehicles regularly send information about their location and driving state towards the server. If a vehicle now needs to predict the behavior of a surrounding object, it requests the prediction from the server. There, a kernel density estimation using all nearby collected measurements is performed. Thus, all location-dependent factors are automatically included in the prediction. Although, the approach seems to produce promising results in a small real-world evaluation, it is questionable whether the solution is also efficiently applicable to a larger area. While doing so, problems in terms of storage capacity as well as computation time could arise. In addition, pure online prediction techniques are suffering from transmission times.

\cite{matute2018longitudinal} uses the curvature of the road as input to plan comfortable trajectories. The plain fact that location-dependent features are also included while planing pleasant trajectories additionally emphasizes that experienced drivers also adapt to changed local conditions.

\cite{qi2014location} studies location-dependent lane change behaviors on arterial roads from the viewpoint of traffic flow analytics. To do so, the authors solve the motion prediction task through classical modelling. The established model distinguishes between efficiency-driven and objective-driven lane changes and aggregates the contributions of both motivations. The efficiency-driven model part on the one hand is primarily based on the motion of surrounding vehicles. The objective-driven part on the other hand mostly relies on the location. To correctly parametrize the model, the authors use the well-studied NGSIM data set \cite{colyar2007us}. Here, the term location-dependent has to be comprehended as microscopic rather than macroscopic. The location-dependent nature of the approach solely refers to the location within the road segment observed in the NGSIM data set, which basically describes in which lane the vehicle is driving and how far away the next intersection is. Macroscopic effects such as differences between regions or cities that our article focuses on are not taken into account.

The article presented in \cite{gonccalves2020change} originates from a completely different community being interested in so-called concept drifts. Concept drifts in general denote changes in classification problems over lifetime. The latter, can make previously well-working classifiers inadequate. \cite{gonccalves2020change} deals with the development of a concept drift detector that in particular is able to detect changes in the a-priori probabilities of the underlying classes. The general idea of a concept drift, even if, spatially than temporally, can also be translated to our maneuver classification problem.

\cite{wang2013learning} also tackles concept drifts in imbalanced online learning applications. The approach dynamically detects changes in the class probabilities and adapts its online learner in order to reflect that.

Concluding our literature study, there exists to the best of our knowledge no research work that spatially aggregates lane change probabilities. This fact may be attributed to the enormous efforts, being necessary to collect the underlying measurement data. Consequently, analyses about location-dependent factors fostering or dampening lane changes have not yet been studied on a large scale.


\section{DATA AND MAP PREPARATION}\label{sec:prep}

\begin{figure}[t!]
\centering\includegraphics[width=0.48\textwidth]{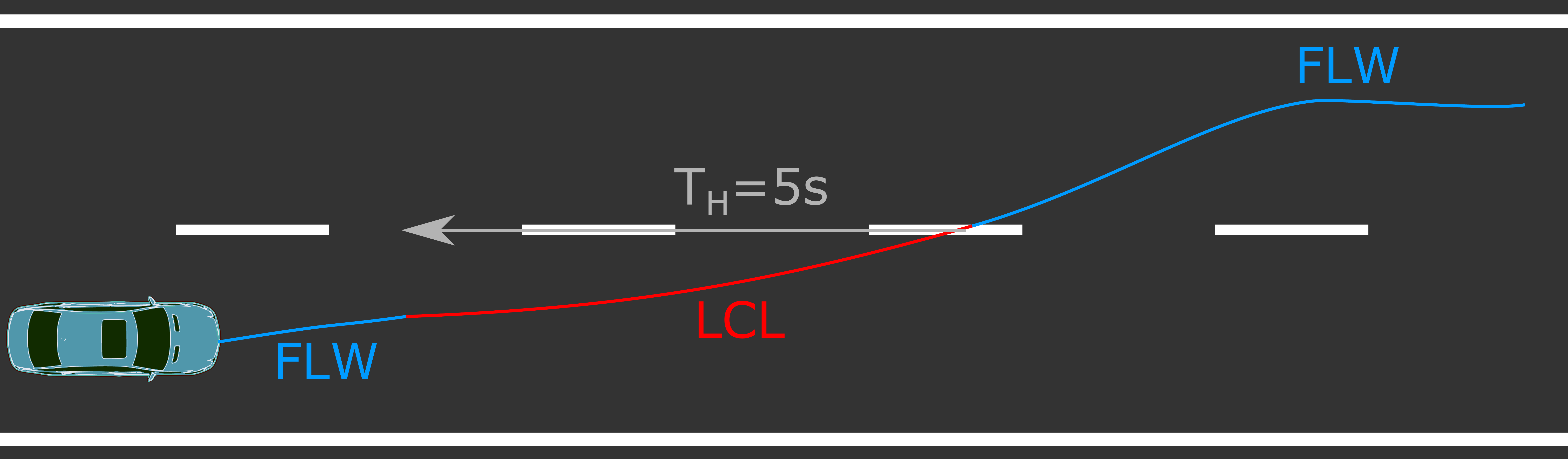}\caption{Simplified visualization of the labeling principle exemplary for a lane change to the left. As shown, all samples 5 seconds prior to the lane crossing are treated as lane change samples, whereas all other samples are treated as lane following ones.}\label{fig:labeling}
\end{figure}

Our empiric investigations build upon data from a large fleet of customer vehicles. All vehicles of our fleet are wirelessly connected with a backend server. From time to time the vehicles send a multitude of selected sensor signals to the backend. Instead of single point measurements for each sensor a time series lasting for up to 200\,s is collected. Prior to data transmission, the communication module anonymizes and encrypts the data. In the backend, all collected measurement data are transformed into an equidistant time-representation with a time resolution of 50\,ms. Besides, only measurements made on German highways are considered. Afterwards, it is possible to detect lane changes carried out in the measurement sequences. Thereby, the continuously measured distances between the vehicle and the lane markings can be used for the detection. These detections can be localized through the GNSS (global navigation satellite system) positions that are sensed as well. For our further investigations we pick the collected measurements of one full day. As a whole, this data set contains data of approximately 1\,350\,h. At this moment, a longer provisioning of data is not possible in order to technical reasons related to the necessity of a massive storage capacity. 

Using the approximately 58\,000 detected lane changes it already becomes possible to do a first visual analysis by arranging them geographically as a heatmap (cf. \autoref{fig:lcmap}). As \autoref{fig:lcmap} already reveals, this investigation however is not very informative, as all larger determined hotspots now comply with congested areas or, more precisely, that areas that are highly frequented by the vehicles of our fleet. Consequently, this result is also not well suited for being used in downstreamed applications.

To overcome this, we assess the lane change events in relation to the amount of lane following events. For this purpose, we assign all measurements to the links of a digital map. Links are one form of representing parts of a digital map and can be considered as a sequence of GNSS points. In addition to its geometry, each link can have several properties as road type, curvature, slope or speed limit. These properties enable further investigations (cf. \autoref{sec:results}). Moreover, we preprocess the map in a way, such that all links have the same length, to be able to compare behavior differences over various links. Furthermore, we shifted from single lane change events to longer measurement parts being categorized as lane changes after some first attempts. Each measurement point being 5 or fewer seconds prior to the crossing of the lane marking is considered as point belonging to that lane change. Note that this is also common practice in the maneuver classification community (e.\,g. in \cite{bahram2016, schlechtriemen2015will, wirthmueller2019}). According to the labeling defined in \cite{wirthmueller2019} each measurement point is therefore assigned to one of the three essential maneuver classes for highway driving: lane change to the left \textit{LCL}, lane following \textit{FLW} and lane change to the right \textit{LCR}. \autoref{fig:labeling} visualizes this labeling principle. Afterwards, it becomes possible to calculate the probabilities of each of the three maneuver classes for each link. Consequently, the lane change behavior at a certain location can be investigated together with all measurements at that position. In particular, using more than a single point per lane change can help to suppress noise biasing the results of later analyses. Note that this is especially important, as lane changes anyway occur by orders of magnitudes rarer than lane following behavior does. Moreover, the probabilities estimated per link can be directly used as local a-priori probabilities for maneuver classification approaches as described in \cite{wirthmueller2019}.


\begin{figure*}[t!]
\centering\includegraphics[width=0.94\textwidth]{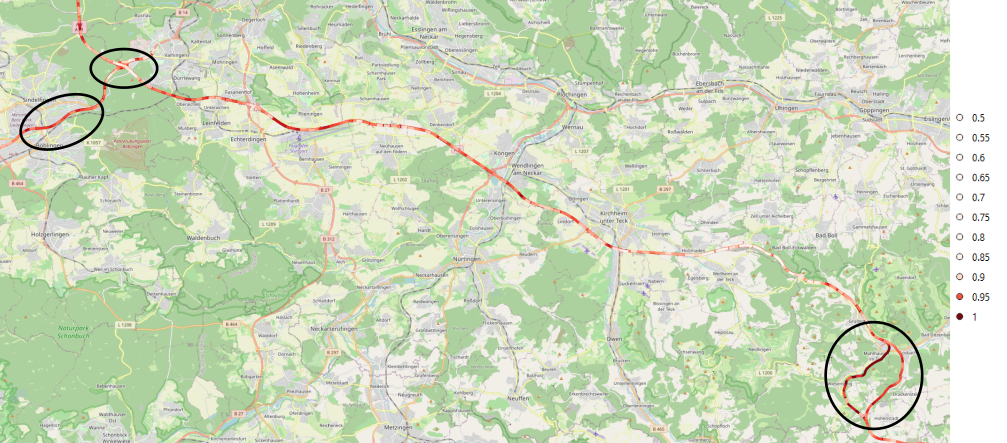}\caption{Visualization of the calculated lane following probability $P_{FLW}$ per 200\,m long link. Probabilities are illustrated in shades of red, where dark red corresponds to a large probability. From a large distance, links being close together might overlap. In addition, the links are visualized according to descending lane following probabilities. Larger areas being colored homogeneously indicate the most interesting locations. Basemap: OpenStreetMap \cite{OpenStreetMap}}\label{fig:globalmap}
\end{figure*}

\begin{figure*}[t!]
\centering\includegraphics[width=0.96\textwidth]{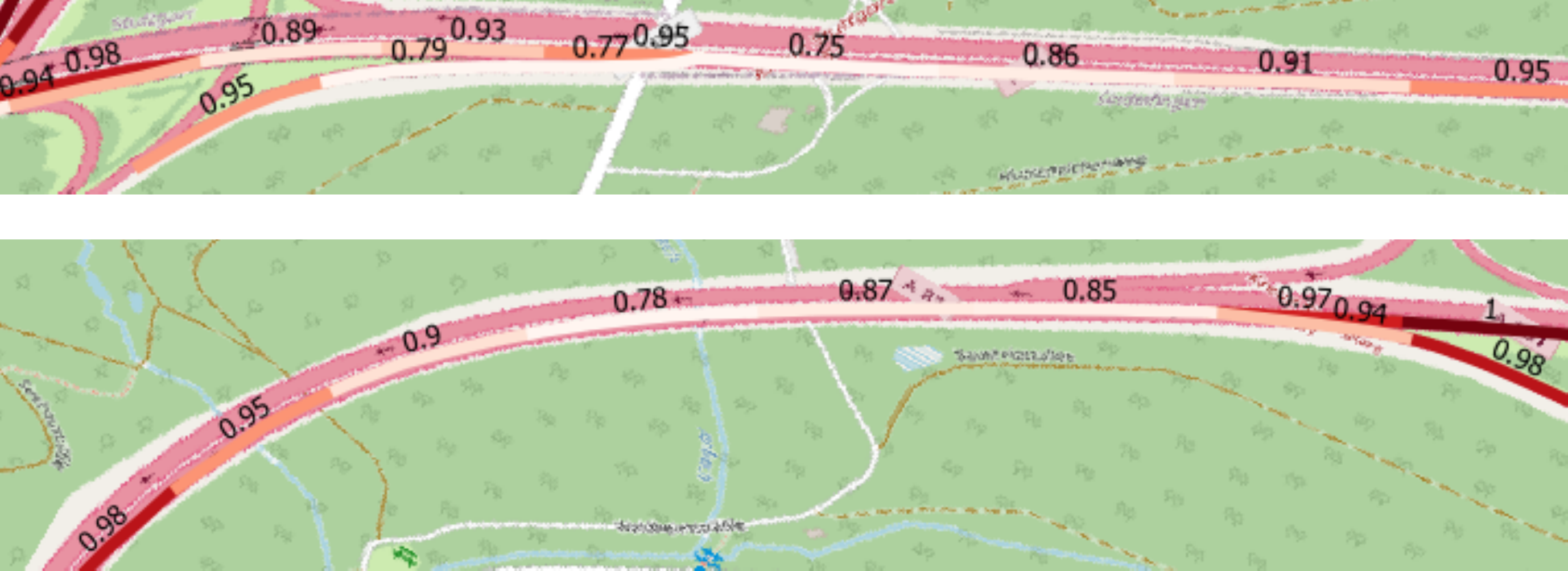}\caption{Microscopic view on the lane following probabilities $P_{FLW}$ per 200\,m long link at a highway merger (upper example) and a highway divider (lower example) respectively at the highway interchange Stuttgart (second left highlighting in \autoref{fig:globalmap}). Basemap: OpenStreetMap \cite{OpenStreetMap}}\label{fig:krstutt}
\end{figure*}

In \cite{wirthmueller2019}, a maneuver classifier is trained using a balanced data set. This means that the proportion of all maneuver classes in the data set is equal, whereas in reality lane changes occur much rarer compared to lane following. Using a balanced data set instead of the unbalanced one enables the machine learning model (in \cite{wirthmueller2019} a multilayer perceptron) to learn the differences between the maneuver classes much easier. Through the extreme over-proportion of lane following events, it would otherwise be favorable for the classifier to assign all samples to that class in order to achieve a good classification performance. However, the class probabilities produced by a classifier trained that way do not correspond to the actual ones. Hence, \cite{wirthmueller2019} suggests weighting the estimated probabilities with the a-priori probabilities of the maneuver classes. At this point it would also be possible to utilize location-dependent rather than global a-priori probabilities. In \autoref{sec:rel_work} we pointed out that pure online prediction approaches are suffering from transmission latencies. In contrast to such approaches location-specific a-priori probabilities are comparably static and can thus be transmitted for larger regions. Thus, short offline phases as well as transmission times do not cause problems.

Note that location-specific a-priori estimates are strongly influenced by the used prediction horizon $T_H$, which we set to 5\,s according to \cite{wirthmueller2019}. A longer horizon results in higher lane change probabilities, whereas a shorter horizon increases the probabilities for lane following. However, the absolute values are not relevant for analyzing location-based differences in the driving behavior. A more important aspect is whether a specific location, geometry or other location-specific properties result in more or less lane changes to one or both neighboring lanes. To get an idea what are more or less lane changes than usual, we again refer to \cite{wirthmueller2019}, where the following overall a-priori probabilities were derived from a large data set: \textit{LCL}: 0.03, \textit{FLW}: 0.94, \textit{LCR}: 0.03.

As most vehicles of our fleet are located in the area around Stuttgart, Germany, we restrict the data to highways in that area in order to base our analyses on a sufficient amount of data per link. Besides, we apply a threshold (i.\,e., 10 measurement points per meter) on the amount of data per link. In some very rare cases, the latter procedure results in few missing links within the considered region. This ensures that the estimated probabilities are less subject to measurement noise. Nevertheless, the data quality can be further improved by increasing the amount of data. This would also help to expand the analytics area. Finally, the resulting data set contains more than 8\,600 lane change maneuvers. 

\section{INVESTIGATIONS AND RESULTS}\label{sec:results}

Already a first look on the map produced after the described preprocessing steps (cf. \autoref{fig:globalmap}) reveals areas which are obviously of special interest. These include highway interchanges, curves and slopes. In the following sections \autoref{subsec:exits} - \autoref{subsec:slopes} we study the respective effects in detail. Besides these identified local conditions causing the effects, there might be even more. As producing a comprehensive collection is, however, not possible in practice we restrict our upcoming investigations to the three given local conditions. \autoref{subsec:discussion} closes the section with an overall discussion of our examinations. 

\subsection{Highway Interchanges}\label{subsec:exits}

A closer look on the lane change probabilities at highway interchanges reveals a non-surprising behavior. \autoref{fig:krstutt} provides examples of more detailed views of highway mergers and dividers. Ahead of dividers and behind mergers, the probability for following the current lane drops. Simultaneously, the probability of lane changes to the right increases ahead of the dividers. This can be traced back to vehicles leaving the current highway. Likewise, approaching vehicles result in increased lane change probabilities to the left behind highway mergers.

\begin{figure*}[t!]
\centering\includegraphics[width=0.98\textwidth]{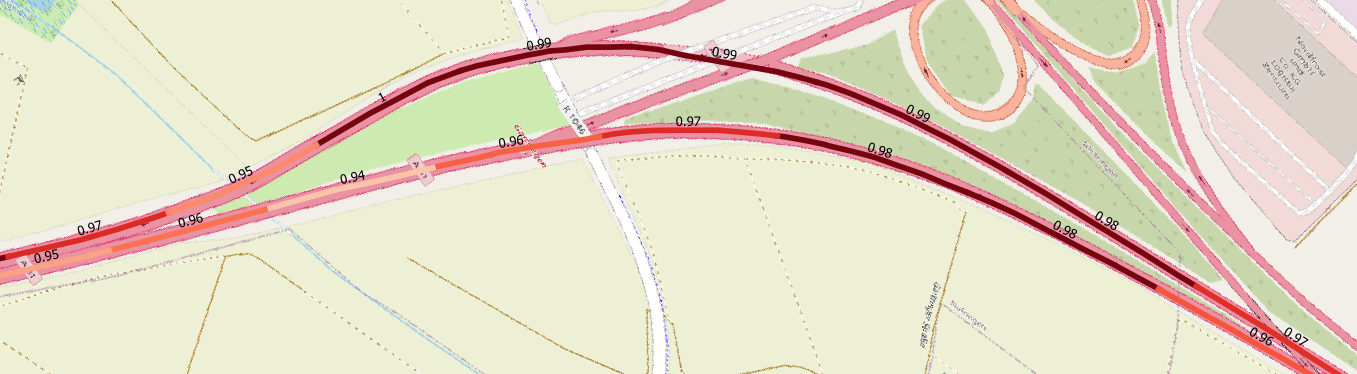}\caption{Microscopic view on the lane following probabilities $P_{FLW}$ per 200\,m long link in a curve close to Gärtringen. Basemap: OpenStreetMap \cite{OpenStreetMap}}\label{fig:curve1}
\end{figure*}

Further, it is noteworthy that in the highway divider example in \autoref{fig:krstutt} the lane change probability to the right (not explicitly visualized) increases already significantly ahead of the actual location of the divider. This effect can be reproduced at some other dividers as well. Apparently, drivers tend to get in lane early. This may be attributed to the specific characteristics of these highway dividers. Possibly, the departing lanes are very long or the signage is set up very early. Note that, such local conditions are very hard to cover with digital maps as well. Besides, the probability to leave the highway can widely vary from one divider to another depending on typical traffic flows. So, it is easy to imagine that each location is very individual.


\begin{figure}[t!]
\centering\includegraphics[width=0.48\textwidth]{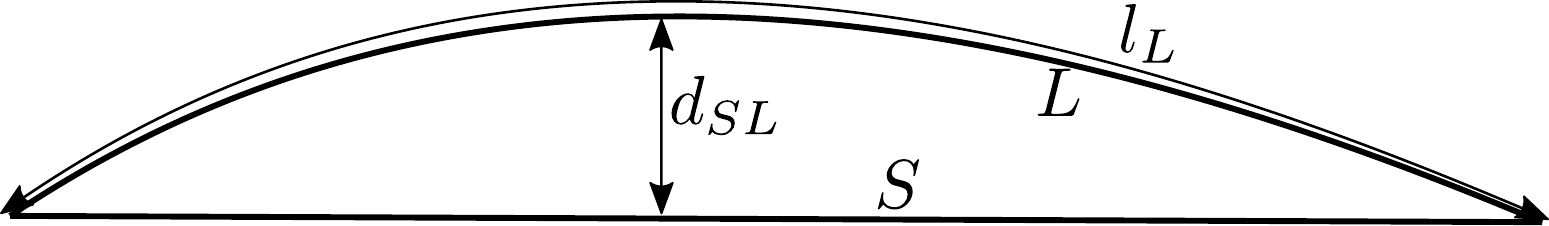}\caption{Calculation of the curvature equivalent $bend$.}\label{fig:bend}
\end{figure} 

\subsection{Curves}\label{subsec:curves}

To enable investigations concerning curves, first, we calculate a new feature called $bend$ for each link. This value is proportional to the average curvature of the link, but can be calculated more efficiently according to \autoref{eq:bend}:

\begin{equation}
bend = \frac{l_L}{d_{SL}}
\label{eq:bend}
\end{equation}

$l_L$ denotes the length of the link $L$. $d_{SL}$ indicates the maximum distance between the link and the secant $S$ connecting its start and end point. \autoref{fig:bend} visualizes this. To distinguish between right and left turns, we select start and end points in such a way that right turns show off negative $bend$ values and left turns positive ones. This ensures that the bend values are in accordance with the vehicle coordinate system specified in the ISO norm 8855 \cite{iso8855} as well.

Note that the $bend$ values are strongly influenced by the used map discretization. For our purpose, a link length of 200\,m has shown the best results. Decreasing the link lengths below 200\,m results in less stable probabilities comprising of more noise. By contrast, links longer than 200\,m are not able to represent curvatures appropriately. Within the selected region, the determined $bend$ values range from -0.33 to 0.14. Obviously, road segments showing especially large bends occur much rarer than slightly curved or straight ones. The largest share of bends ranges between $\pm$0.07.

\begin{figure}[t!]
\centering\includegraphics[width=0.48\textwidth]{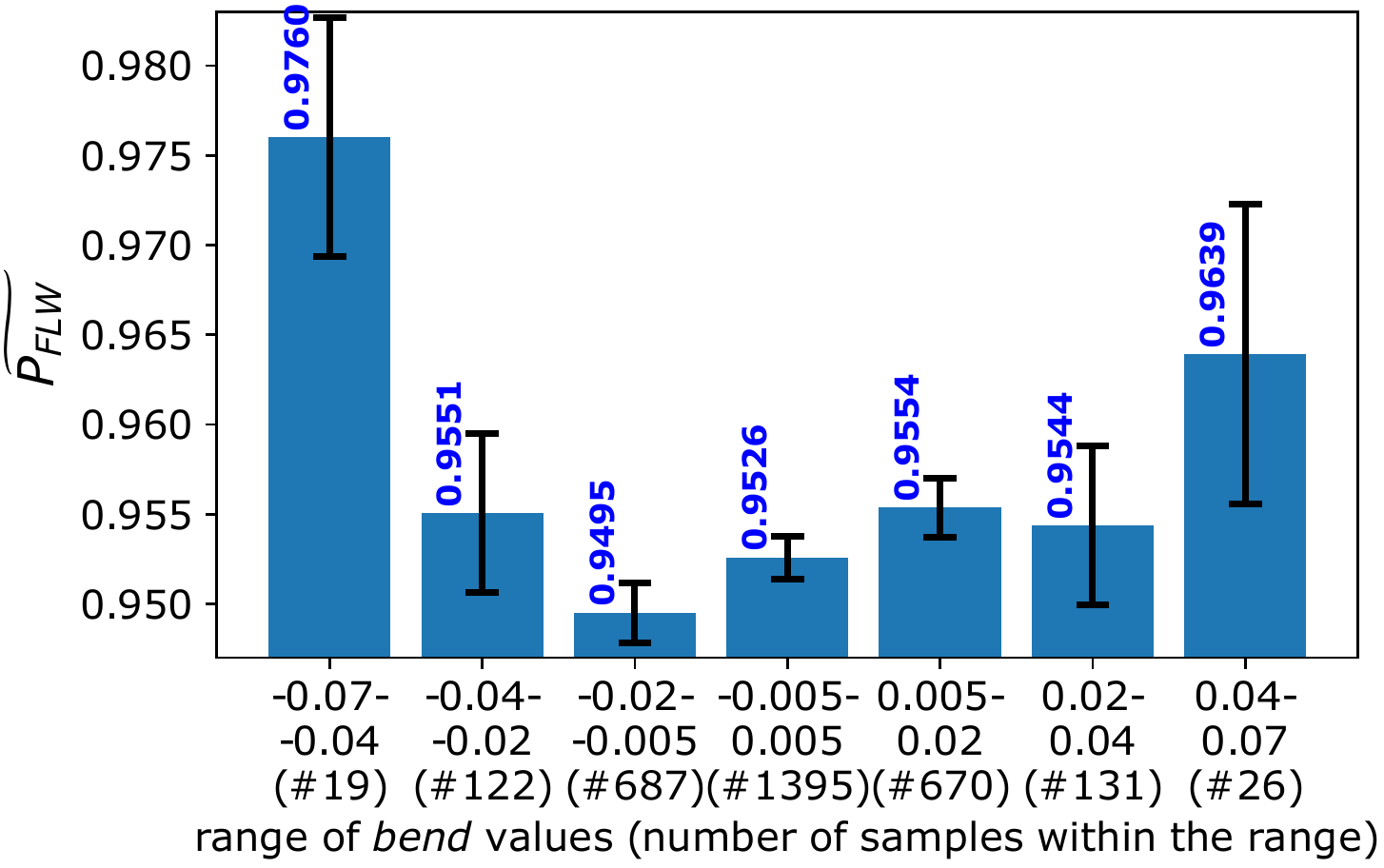}\caption{Median lane following probability under certain $bend$ values of 200\,m long links. Error bars denote standard error of the measurement (SEM).}\label{fig:bend_vs_prob}
\end{figure}

In order to quantify the impact of curves on the lane change behavior \autoref{fig:bend_vs_prob}, shows off the median probability for lane following $\widetilde{P_{FLW}}$ at links within certain $bend$ ranges. Road segments with very large $bends$ ($|.|$\,$>$\,$0.07$) were excluded due to the low data densities and resulting noise level. From \autoref{fig:bend_vs_prob} one can conclude that larger bends result in larger lane following probabilities. Accordingly, curves seem to discourage drivers from changing lanes. However, note that the observable differences are rather small considering absolute values.

\footnotetext[1]{sign depends on the driving direction}

Taking a look at some specific locations substantiates the existence of the described effect. \autoref{fig:curve1} shows an example of a rather strongly curved road segment with $bend$ values ranging from $\pm$0.045\footnotemark[1] to $\pm$0.071\footnotemark[1] in the links belonging to the curve. The annotated lane following probabilities indicate that drivers tend to perform lane changes ahead of or behind the respective curve. Even though it has to be mentioned that close to that curve a particularity, i.\,e. one of just a few left-hand highway exits in Germany, is located. This should, however, make lane changes even more probable.

\begin{figure*}[t!]
\centering\includegraphics[width=0.98\textwidth]{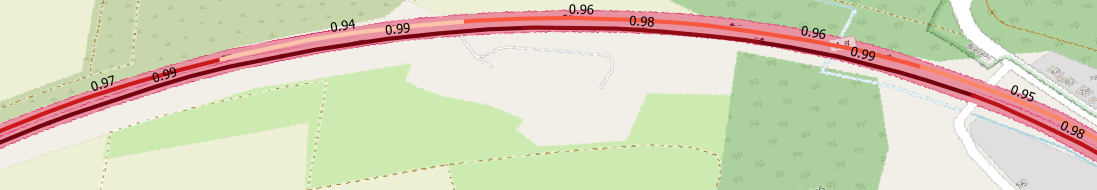}\caption{Microscopic view on the lane following probabilities $P_{FLW}$ per 200\,m long link in a curve close to Pforzheim. Basemap: OpenStreetMap \cite{OpenStreetMap}}\label{fig:curve2}
\end{figure*}

\autoref{fig:curve2} shows another microscopic view of a curve, where the lane change-dampening impact can be observed especially if the curve is driven through as a right turn. Within the shown curve, the $bend$ values are nearly constant around $\pm$0.030\footnotemark[2]. Note that the probability of lane changes to the left seems to decrease particularly within the right turn. Probably this can be explained by the fact that drivers who perform lane changes to the left in a right turn may get the feeling of being pushed out of the curve. Another possible explanation could be the reduced visibility of vehicles on the outer adjacent lane.

In addition, other publications \cite{wirthmueller2019, dang2017time} argue that lane changes to the right are mostly motivated by the intention to leave the highway (objective-driven cf. \cite{qi2014location}), whereas lane changes to the left are mostly attributed to the intention of overtaking slower leading vehicles (efficiency-driven cf. \cite{qi2014location}). This means lane changes to the right might be more necessary to reach the destination. In turn, this might explain why the effect seems to be stronger here compared to lane changes to the left.

Further note that the latter effect complies with another conclusion one may draw from \autoref{fig:bend_vs_prob}. Comparing the two bars corresponding to the strongest right curves (leftmost bar) and left curves (rightmost bar), reveals that right curves seem to have a stronger lane change-dampening effect than left curves.

In conclusion, many indications seem to support our intuition that curves take a lane change-dampening effect. Especially, the superposition of various effects, however, makes it difficult to prove this.

\begin{figure}[t!]
\centering\includegraphics[width=0.48\textwidth]{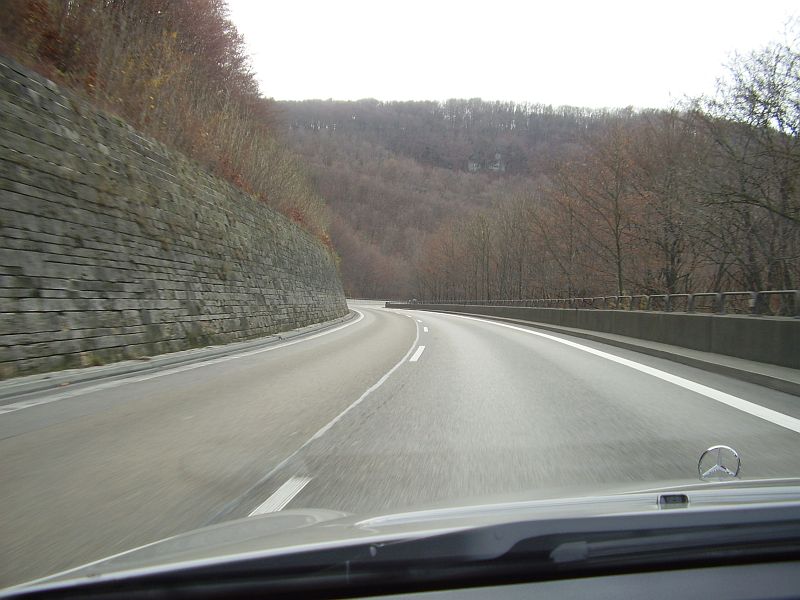}\caption{Picture taken at the leading up part of the road at `Drackensteiner Hang'. Image source:\href{http://www.autobahn-bilder.de/inlines/a8.htm}{http://www.autobahn-bilder.de/inlines/a8.htm}}\label{fig:drackenstein}
\end{figure}

\footnotetext[2]{sign depends on the driving direction}

\subsection{Slopes}\label{subsec:slopes}
Slopes constitute another spatial particularity being of interest in the context of lane change behavior. Compared to curves and highway interchanges, however, slopes cannot be easily identified on a navigation map, instead one has to rely on expert or local knowledge to identify such correlations at all.

Our attention to this potential correlation was drawn while taking a look at the rightmost highlighting in \autoref{fig:globalmap}. This location is known as `Albaufstieg' at the so-called `Drackensteiner Hang' \cite{Wikipedia}. The latter is an approximately 6\,km long inclination with a quiet constant slope around 5\,\%. Due to the particular topology of this area the roads for the two directions are not running in parallel. \autoref{fig:drackenstein} shows a picture taken at the leading up part of that road. As can be seen, the road course at this location is not only very steep, but is also hard to overlook due to the additional curvature and narrowing road limits. Our lane change investigation reveals that exactly at that location hardly any lane changes have been performed. 

The corresponding down hill part of the road, in turn shows off only a slightly reduced lane change probability. In summary, this special location supports our impression that slopes and especially uphill parts could have a similar lane change-dampening effect as curves.

\begin{figure}[t!]
\centering\includegraphics[width=0.48\textwidth]{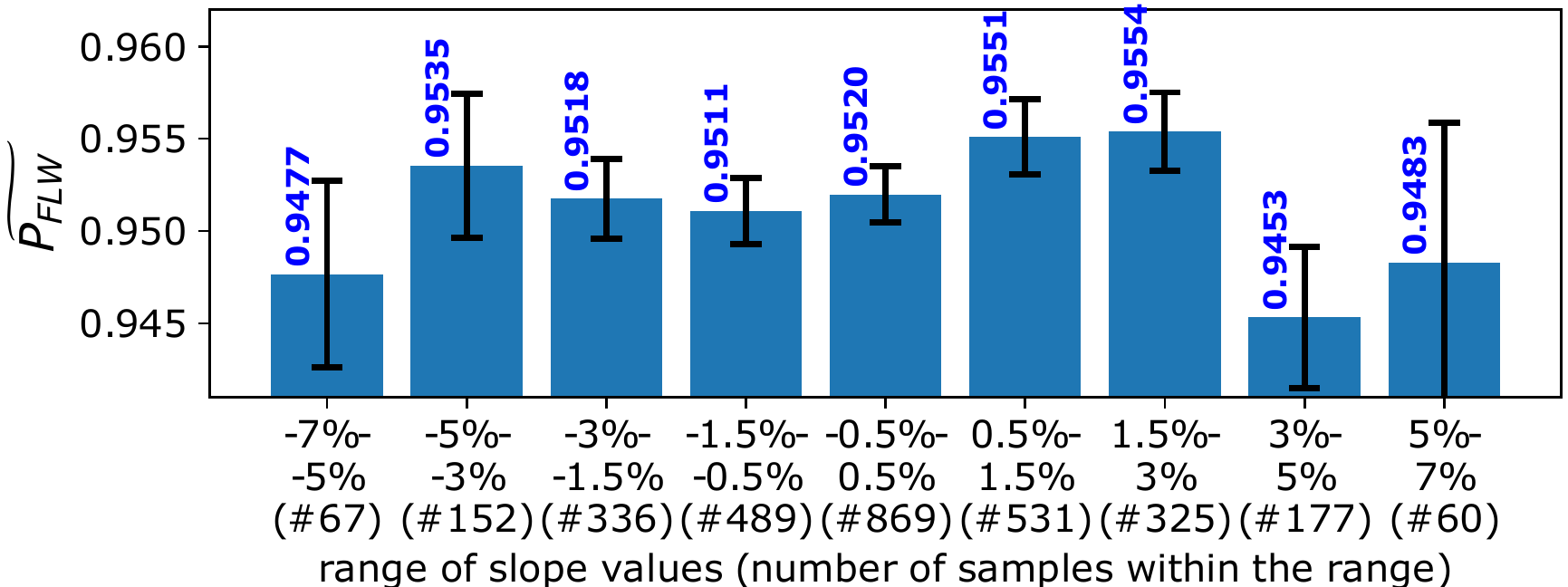}\caption{Median lane following probability under certain $slope$ values of 200\,m long links. Error bars denote standard error of the measurement (SEM).}\label{fig:slope_vs_prob}
\end{figure}

\begin{figure*}[t!]
\centering\includegraphics[width=0.93\textwidth]{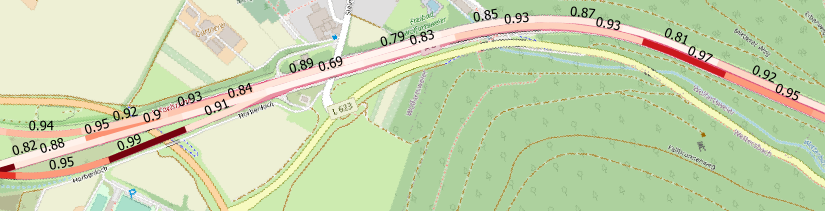}\caption{Microscopic view on the lane following probabilities $P_{FLW}$ per 200\,m long link at a location close to Karlsruhe.}\label{fig:karlsruhe}
\end{figure*}

To investigate this in more detail, \autoref{fig:slope_vs_prob} shows the median lane following probability $\widetilde{P_{FLW}}$ at links (with 200\,m length) under certain $slope$ values. The illustration is equivalent to that for the curvatures in \autoref{fig:bend_vs_prob}. The shown slopes are also in accordance with the vehicle coordinate system defined by ISO norm 8855 \cite{iso8855}. Positive slope values indicate ascents, whereas negative values indicate descents. Most slopes on highways are ranging between $\pm$7\,\%. To the best of our knowledge, steeper roads constitute an absolute exception, at least on highways in Central Europe. Already values around 5\,\% enduring over several kilometers as in the case of the `Drackensteiner Hang' are very rare.

Taking a more detailed view on \autoref{fig:slope_vs_prob} reveals that it is not possible to detect an unambiguous trend fostering our intuition that slopes have a lane change-dampening effect. Explanations for this phenomenon may be manifold. Obviously, once more, several effects probably superimpose each other. A good example of such a superposition can be found at a location close to Karlsruhe, shown in \autoref{fig:karlsruhe}. At this location the lane change probabilities are increased even though the road course is rather steep with slope values ranging from $\pm$4.5\,\%\footnotemark[3] to $\pm$7.3\,\%\footnotemark[3]. These lane changes are, however, attributed to the close highway interchange. Solely removing the links at this location from the examination in \autoref{fig:slope_vs_prob} takes a significant effect. The latter corrects the lane following probabilities in the bins corresponding to the largest slopes notably upwards.

In summary, our overall impression is that slopes can contribute in a lane change-dampening manner but solely in combination with other conditions. In this way, slopes can, for instance, in combination with curves make locations hard to overlook, whereby the willingness for lane changes can be decreased. On the other hand, slopes can also lead to opposite effects at other locations, e.\,g. if poorly motorized vehicles and trucks are slowed down by ascents. This could provide additional motivation to other traffic participants to change lanes in order to overtake the slower vehicles. 

\subsection{Discussion}\label{subsec:discussion}

The presented examinations clearly emphasize that location-dependent effects on the lane change behavior definitely exist but are hard to investigate and quantify. Of course this is also attributed to the superposition of multiple effects. Nevertheless, the investigations have revealed the following tendencies:

\begin{itemize}
\item As to be expected, highway mergers and dividers can foster the likelihood of lane changes.
\item Curved road segments can have lane change-dampening effects.
\item Slopes can affect lane change probabilities in both directions depending on other simultaneous conditions. 
\end{itemize}

However, it is extremely challenging to identify which of the considered conditions are important at a certain location and how far they superimpose each other. Moreover, there are obviously much more than the three described factors that impact lane change probabilities. For instance, the time of day, whether it is weekend or a work day, current weather conditions or even season-dependent position of the sun have the potential to influence lane change probabilities.

So on the one hand generalizing in order to predict lane change probabilities based on environment conditions is obviously a very hard task as there are so many factors and interdependencies. On the other hand, we actually do not see the necessity to employ such a generalized model. At least in Germany the highway road network is comparably stable and with an overall length of about 13\,000\,km rather manageable. Instead of error-prone and complex generalizations we suggest building up a lane change probability map containing lane change probabilities for all map links. This straight forward enumeration becomes possible based on data from a broad customer fleet. New `unseen' highway links will quickly be added and updated to the atlas of lane change probabilities.



 \footnotetext[3]{sign depends on the driving direction}

\section{SUMMARY AND OUTLOOK}\label{sec:conclusion}

This article showed how to develop a lane change probability map using measurement data collected with a fleet of customer vehicles. These data are spatially aggregated using digital maps. This enabled analyses revealing the impact of three exemplary location-dependent conditions on the lane change behavior. More precisely, the effects of highway interchanges, curvatures and slopes have been thoroughly investigated and discussed. Although the results show clear tendencies, they nevertheless demonstrate that especially the superposition of several factors makes it hard or even impossible to estimate location-specific lane change probabilities using simple models. Thus, we suggest dynamically constructing and maintaining such a lane change map.

Future work will focus on providing and integrating the information of such a lane change probability map to enhance onboard predictions of surrounding traffic participants behavior. For this purpose, it is also reasonable to collect data over longer time horizons. We are currently integrating our processing chain into a streaming layer. Thus, only final results rather than raw sensor measurements will be provisioned. This will solve our described difficulties concerning data provisioning. The latter will enable us to increase our data basis by an order of magnitudes as well as to enlarge the analytics area. Thus, more advanced analytics and statistical tests become possible, as this increases, for example, the amount of strongly curved or ascending road segments.


\section*{Acknowledgment}

Map data copyrighted OpenStreetMap contributors and available from \href{https://www.openstreetmap.org}{https://www.openstreetmap.org}


\balance

\bibliographystyle{ieeetr}

\bibliography{bib_itsc2021}

\interlinepenalty=10000
\begin{thebibliography}{10}

\bibitem{OpenStreetMap}
{OpenStreetMap contributors}, ``{Planet dump retrieved from
  https://planet.osm.org }.'' \url{ https://www.openstreetmap.org }, 2017.

\bibitem{lefevre2014}
S.~Lef{\`e}vre, D.~Vasquez, and C.~Laugier, ``A survey on motion prediction and
  risk assessment for intelligent vehicles,'' {\em ROBOMECH journal}, vol.~1,
  no.~1, p.~1.
\newblock Nature Publishing Group, 2014.

\bibitem{mozaffari2020deep}
S.~Mozaffari, O.~Y. Al-Jarrah, M.~Dianati, P.~Jennings, and A.~Mouzakitis,
  ``Deep learning-based vehicle behavior prediction for autonomous driving
  applications: A review,'' {\em IEEE Transactions on Intelligent
  Transportation Systems (T-ITS)}.
\newblock IEEE, 2020.

\bibitem{wirthmueller2020}
F.~Wirthm{\"u}ller, J.~Schlechtriemen, J.~Hipp, and M.~Reichert, ``Towards
  incorporating contextual knowledge into the prediction of driving behavior,''
  in {\em 23th International Conference on Intelligent Transportation Systems
  (ITSC)}, IEEE, 2020.

\bibitem{wirthmuller2020fleet}
F.~Wirthm{\"u}ller, M.~Klimke, J.~Schlechtriemen, J.~Hipp, and M.~Reichert, ``A
  fleet learning architecture for enhanced behavior predictions during
  challenging external conditions,'' in {\em 2020 IEEE Symposium Series on
  Computational Intelligence (SSCI)}, pp.~2739--2745, IEEE, 2020.

\bibitem{imanishi2020model}
Y.~Imanishi, Y.~Iihoshi, Y.~Okuda, and T.~Okada, ``Model-less location-based
  vehicle behavior prediction for intelligent vehicle,'' in {\em 2020 IEEE
  Intelligent Vehicles Symposium (IV)}, pp.~716--722, IEEE, 2020.

\bibitem{matute2018longitudinal}
J.~A. Matute, M.~Marcano, A.~Zubizarreta, and J.~Perez, ``Longitudinal model
  predictive control with comfortable speed planner,'' in {\em 2018 IEEE
  International Conference on Autonomous Robot Systems and Competitions
  (ICARSC)}, pp.~60--64, IEEE, 2018.

\bibitem{qi2014location}
H.~Qi, D.~Wang, P.~Chen, and Y.~Bie, ``Location-dependent lane-changing
  behavior for arterial road traffic,'' {\em Networks and Spatial Economics},
  vol.~14, no.~1, pp.~67--89.
\newblock Springer, 2014.

\bibitem{colyar2007us}
J.~Colyar and J.~Halkias, ``{US} highway 101 dataset,'' {\em Federal Highway
  Administration (FHWA), Tech. Rep. FHWA-HRT-07-030}, 2007.

\bibitem{gonccalves2020change}
P.~M. Gon{\c{c}}alves, R.~S. Barros, and S.~Chartier, ``A change detector for
  prior probabilities of classes,'' in {\em 2020 IEEE Symposium Series on
  Computational Intelligence (SSCI)}, pp.~1029--1036, IEEE, 2020.

\bibitem{wang2013learning}
S.~Wang, L.~L. Minku, and X.~Yao, ``A learning framework for online class
  imbalance learning,'' in {\em 2013 IEEE Symposium on Computational
  Intelligence and Ensemble Learning (CIEL)}, pp.~36--45, IEEE, 2013.

\bibitem{bahram2016}
M.~Bahram, C.~Hubmann, A.~Lawitzky, M.~Aeberhard, and D.~Wollherr, ``A combined
  model-and learning-based framework for interaction-aware maneuver
  prediction,'' {\em IEEE Transactions on Intelligent Transportation Systems
  (T-ITS)}, vol.~17, no.~6, pp.~1538--1550.
\newblock IEEE, 2016.

\bibitem{schlechtriemen2015will}
J.~Schlechtriemen, F.~Wirthmueller, A.~Wedel, G.~Breuel, and K.-D. Kuhnert,
  ``When will it change the lane? {A} probabilistic regression approach for
  rarely occurring events,'' in {\em 14th Intelligent Vehicles Symposium (IV)},
  pp.~1373--1379, IEEE, 2015.

\bibitem{wirthmueller2019}
F.~Wirthm{\"u}ller, J.~Schlechtriemen, J.~Hipp, and M.~Reichert, ``Teaching
  vehicles to anticipate: A systematic study on probabilistic behavior
  prediction using large data sets,'' {\em Transactions on Intelligent
  Transportation Systems (T-ITS)}.
\newblock IEEE, 2020.

\bibitem{iso8855}
{DIN ISO 8855}, ``Stra{\ss}enfahrzeuge, {F}ahrzeugdynamik und
  {F}ahrverhalten,'' 2012.

\bibitem{dang2017time}
H.~Q. Dang, J.~F{\"u}rnkranz, A.~Biedermann, and M.~Hoepfl,
  ``Time-to-lane-change prediction with deep learning,'' in {\em 2017 IEEE 20th
  International Conference on Intelligent Transportation Systems (ITSC)},
  pp.~1--7, IEEE, 2017.

\bibitem{Wikipedia}
{Wikipedia contributors}, ``Wikipedia: Drackensteiner {H}ang.''
  \url{https://en.wikipedia.org/wiki/Drackensteiner_Hang}.
\newblock Accessed: 2021-03-29.

\end{thebibliography}

\end{document}